\documentclass[11pt,letterpaper]{article}

\usepackage[utf8]{inputenc}

\usepackage{latexsym}
\usepackage{verbatim}
\usepackage[a4paper, total={6.5in, 9in}]{geometry}
\usepackage{silence}
\usepackage{amsthm,amsmath,amsfonts,amssymb,verbatim, color}
\usepackage{graphicx}
\usepackage[percent]{overpic}
\usepackage{bm}
\usepackage{braket}
\usepackage{float}
\usepackage{mathtools}
\usepackage{appendix}
\usepackage[colorlinks=true,citecolor=blue,linkcolor=black,urlcolor=blue]{hyperref}
\usepackage{graphics} 
\graphicspath{ {figures/} }
\usepackage[caption=false]{subfig}

\usepackage[mathscr]{euscript}
\usepackage[export]{adjustbox}
\usepackage[bottom]{footmisc}
\usepackage{textcomp}
\usepackage{braket}

\def\mA{\mathcal A}
\def\mE{\mathcal E}

\begin{document}

\title{\bf Qubit control using quantum Zeno effect: Action principle approach }
\author{$^{1,2}$Komal Kumari, $^{1}$Garima Rajpoot, $^{1,4}$Sandeep Joshi, and $^{1,2,3}$Sudhir Ranjan Jain \\ $^{1}${\small{\it Theoretical Nuclear Physics and Quantum Computing Section}}\\ {\small{\it Nuclear Physics Division, Bhabha Atomic Research Centre, Mumbai 400085, India}}\\
$^{2}${\small{\it Homi Bhabha National Institute, Training School Complex, Anushakti Nagar, Mumbai 400094, India}}\\
$^{3}${\small{\it UM-DAE Centre for Excellence in Basic Sciences, University of Mumbai}}\\ {\small{\it Vidyanagari campus, Mumbai 400098, India}}\\
$^{4}${\small{\it Department of Physics and Astronomy, Northwestern University}}\\ {\small{\it Evanston, Illinois, U.S.A.}}
}

\maketitle
\begin{abstract}
We employ the stochastic path-integral formalism and action principle for continuous quantum measurements - the Chantasri-Dressel-Jordan (CDJ) action formalism \cite{jordan,jordan2} - to understand the stages in which quantum Zeno effect helps control the states of a simple quantum system. The detailed dynamics of a driven two-level system subjected to repeated measurements unravels a myriad of phases, so to say. When the detection frequency is smaller than the Rabi frequency, the oscillations slow down, eventually coming to a halt at an interesting resonance when measurements are spaced exactly by the time of transition between the two states. On the other hand, in the limit of large number of repeated measurements, the dynamics organizes itself in a rather interesting way about two hyperbolic points in phase space whose stable and unstable directions are reversed. Thus, the phase space flow occurs from one hyperbolic point to another, in different ways organized around the separatrices. We believe that the systematic treatment presented here paves the way for a better and clearer understanding of quantum Zeno effect in the context of quantum error correction.      
\end{abstract}

\section{Introduction}
Bohr's quantum jump is accompanied by the absorption or emission of a photon in case of a radiative resonant energy transfer. In a quantum system there are  continuous transitions, making it difficult to monitor or control them. To monitor individual transitions, Dehmelt \cite{dehmelt,deh} proposed a scheme wherein, in addition to the two levels of the system, there is also a third metastable state which is much longer lived. So the system undergoes transitions between the ground and upper levels, but at a certain random moment, it might transit to the metastable state and remain shelved there, until finally it returns to the ground state. This idea has been beautifully implemented by Minev et al. \cite{minev} where instead of an atom, an artificial atom has been created  by hybridizing transmons. Here the role of metastable state is played by a dark state which is driven weakly and is not being observed. The excited state of the two-level system is continuously monitored by a dispersive coupling to the cavity. When the shift in frequency occurs which is different from what is expected from Rabi oscillations, a jump into the dark state is registered. This physical situation may also be realized by considering a two-level system whose excited state is being measured repeatedly, thus interrupting the Rabi oscillations. QZE is a result of repeated projective, instantaneous measurements on a system whereby the evolution is frozen \cite{ms}. Projective measurements assume performing instantaneous ideal measurements where the response time of the measuring apparatus is much shorter than other relevant time scales \cite{ks}. The intrinsic quantum fluctuations of the detector brings stochasticity to quantum evolution, leading to a description in terms of quantum trajectories. Going beyond the projective measurements, the quantum Zeno regime has a rich structure as a function of frequency of measurement. The ``cascades of transition" brought out recently \cite{parveen} appear as ``phases of QZE" when a quantum system is subjected to weak continuous measurements. 

For practical reasons and in order to address a larger class of phenomena, we need to consider a wider class of measurements where only partial information about an observable may be extracted. In contrast with projective measurements, there exist many measurement models where measured values of an observable of interest possess some  uncertainty, howsoever small \cite{jacobs-steck}. One such method is subsumed in what is known as diffusive measurement. The two outcomes (say, 0 and 1) are discerned by the detection system as probability distributions about possible values of some physical quantity (current, intensity, etc.). These occur in a certain range, modelled by the standard deviation of the distribution to which the values belong. These Poisson distributed uncorrelated outcomes,  by the law of large numbers \cite{kac}, lead to a Gaussian distribution for measured quantities. As we shall see in our discussion on diffusive measurements, the appropriate measurement operators are constructed on this basis. 

Action has a very special place in physics, Ehrenfest showed that the quantization of energy levels is connected to the adiabatic invariance of classical action \cite{ehrenfest}. Since a change in the value of adiabatic invariant is in terms of jumps \cite{crawford,jain2005}, action changes discontinuously upon repeated measurements. Interestingly, in Sec. 3.2.2, we bring out the validity of this idea. 

If a quantum system is interacting with environment, its evolution can be described as a solution of a stochastic master equation where the environment needs to be monitored. The solution consists of a multitude of paths having different weights, constituting the quantum trajectories \cite{carmichael}. Quantum trajectory describes the conditional state of knowledge of a system, given its measurement record.  This evolution can only be considered `deterministic' when the measurement record is fixed, independent of discrete or continuous measurements. We calculate the most optimal path and jump events in a phase space representation. Working with individual trajectories and jumps is expected to help correcting errors in a quantum system whereby  allowing us to monitor and manipulate the jumps mid-flight \cite{minev}. Quantum Zeno effect, which is at the heart of this adaptation of Dehmelt's idea in superconducting qubits, has been shown to possess a nontrivial ``cascade of transitions" \cite{parveen}. In this work, we employ the action principle developed in \cite{jordan,jordan2} to systematically address quantum Zeno effect and quantum jumps by resorting to clear depictions in phase space. In phase space representation, it turns out that the transition points are, in fact, saddle points. Interestingly, the pair of expanding and contracting directions in phase space are inverted for the two points. There also appear a pair of separatrices. Altogether, the phase space is divided by these special structures, guiding the phase flow and the corresponding transitions and jumps in a rather intricate and instructive way.  

\section{The CDJ formalism for quantum measurements}\label{CDJ formalism}

Let us say that we have performed a series of measurements on a two-level system for total time $T$, entailing a realization of a measurement, $\{r_k\}_{k=0}^{n-1}$ \cite{jordan,jordan2}. Each readout is obtained between time $t_k$ and $t_{k+1}=t_k+\delta t$. Define a series of quantum states $\{\textbf{q}(t_k)\}_{k=0}^n$ at the series of time $\{t_k\}_{k=0}^n$, written as a $d$-dimensional parametrized vector $\textbf{q}$, where the components are coefficients of expansion of the evolution of density operator, $\rho$, written in some orthogonal operator basis, such as the Pauli $\sigma$ basis of a two-state system. For this system, $\textbf{q}=\{x,y,z\}$ are the coordinates of the Bloch sphere.

The quantum trajectory can be computed via an update equation of the form, $\textbf{q}(t_{k+1})={\mathcal O}[\textbf{q}(t_{k}),r_k]$, which includes the back-action from the measurement readout and the unitary evolution from system Hamiltonian. In the Markovian approach, we can write the Joint Probability Distribution Function (JPDF) of all measurement outcomes and state trajectories, which gives all the statistical information about a system. The JPDF can be written as: 
\begin{alignat}{1}\label{eq:jpdf}
    \mathcal{P}_\zeta&\equiv P(\{\textbf{q}(t_{k}))\}_1^n,\{r_k\}_0^{n-1}|\textbf{q}_0,\zeta),\nonumber\\
    &=B_\zeta\prod_{k=0}^{n-1}P(\textbf{q}(t_{k+1})|\textbf{q}(t_{k}),r_k)P(r_k|\textbf{q}(t_{k})),
\end{alignat}
where $P(r_k|\textbf{q}(t_{k}))$ is the conditional probability distribution for the measurement outcome $r_k$, given the state of the system before the measurement is $\textbf{q}(t_{k})$. Moreover, $P(\textbf{q}(t_{k+1})|\textbf{q}(t_{k}),r_k)=\delta^d(\textbf{q}(t_{k+1})-{\mathcal O}[\textbf{q}(t_{k}),r_k])$ is the deterministic conditional probability distribution for the quantum state after the measurement, conditioned on the state at the previous time step and measurement readout. $B_\zeta=B_\zeta[\{\textbf{q}(t_{k})\},\{r_k\}]$ is a function accounting for subensembles, i.e., where an initial or final or both states have been chosen.

For an arbitrary functional, $\mathcal{A}=\mathcal{A}[\{\textbf{q}(t_{k})\},\{r_k\}]$, its expectation value is given by the functional integral, $\langle\mathcal{A}\rangle_\zeta=\int d[\textbf{q}(t_{k})]_1^n d[r_k]_0^{n-1}\mathcal{P}_\zeta \mathcal{A}$, where $\int d[\textbf{q}(t_{k})]_1^n\equiv\int d\textbf{q}(t_1)\dots d\textbf{q}(t_n)$. Direct integration of $\mathcal{A}$ will be tedious even for a single qubit measurement problem. Following \cite{jordan}, we write the JPDF as a path integral with a suitably defined action functional.

We write $\delta^d(\textbf{q}(t_{k+1})-{\mathcal O}[\textbf{q}(t_{k}),r_k])$ for $k=0$ to $n-1$ in the Fourier integral form:
\begin{equation}
    \delta(q)=\frac{1}{2\pi\iota}\int_{-\iota\infty}^{\iota\infty}e^{-pq}dp,
\end{equation}
for each component of $\textbf{q}$ and rewrite other terms in an exponential form. The conjugate variables for $\delta$-functions are denoted by $p(t_{k})$, $k=0$ to $n-1$. The final form of JPDF is then
\begin{equation}
    \mathcal{P}_\zeta=\mathcal{N}\int d[\textbf{p}(t_{k})]\exp(\mathcal{S})=\mathcal{N}\int\mathcal{D}\textbf{p}\exp(\mathcal{S}),
\end{equation}
in the limit $\delta t\to 0$, where for functional integrals, $\int\mathcal{D}\textbf{p}\equiv\lim_{\delta t\to 0}\int d[\textbf{p}(t_{k})]$and $\mathcal{N}$ is the normalization factor.

The action is then given by,
\begin{equation}\label{eq:action}
    \mathcal{S}(\textbf{p},\textbf{q},r)=\mathcal{B}_\zeta+\int_0^T\delta t\{-\textbf{p}\cdot(\dot{\textbf{q}}-\mathcal{L}[\textbf{q},r])+\mathcal{F}[\textbf{q},r])\}
\end{equation}
where we have introduced $\dot{\textbf{q}}\delta t=\mathcal{L}[\textbf{q},r]\delta t$ as the continuous time version of the state-update equation $\textbf{q}(t_{k+1})={\mathcal O}[\textbf{q}(t_{k}),r_k]$. This update equation comes from the state transformation equation,
\begin{equation}\label{eq:density}
    \hat{\rho}(t+\delta t)=\frac{\mathcal{M}\mathcal{U}\hat{\rho}(t)\mathcal{U}^\dagger \mathcal{M}^\dagger} {Tr[\mathcal{M}\mathcal{U}\hat{\rho}(t)\mathcal{U}^\dagger\mathcal{M}^\dagger]},
\end{equation}
where $\mathcal{M}$ is the evolution operator resulting in measurement back-action. An expansion in powers of $\delta t$ entails $P(r_k|\textbf{q}(t_{k}))\propto\exp\{\delta t\mathcal{F}[\textbf{q}(t_{k}),r_k]+\mathcal{O}(\delta t^2)\}$; we define  $\mathcal{F}[\textbf{q}(t_{k}),r_k]=\ln{P(r_k|\textbf{q}(t_{k}))}$. The Hamiltonian is,
\begin{alignat}{1}\label{eq:ham}
    \mathcal{H}(\textbf{p},\textbf{q},r)&=\textbf{p}\cdot\mathcal{L}[\textbf{q},r]+\mathcal{F}[\textbf{q},r]-\textbf{p}\cdot(\textbf{q}-\textbf{q}_I)\delta (t)-\textbf{p}\cdot(\textbf{q}-\textbf{q}_F)\delta(t-T)
\end{alignat}
for continuum limit, obtained by taking $\delta t \to 0$, $n\to\infty$ and setting $t_0=0$, $t_n=T$ for a subensemble with initial and final boundary conditions to the states, $\textbf{q}(t_0)=\textbf{q}(t_I)$ and $\textbf{q}(t_n)=\textbf{q}(t_F)$.

We can find the largest contribution of path integral by extremizing the action. Taking the first variation of action over all variables and setting $\delta S$ to $0$ in \eqref{eq:action}, appealing to the principle of extremum action, we have,
\begin{alignat}{1}\label{eq:optimal}
    -\dot{\textbf{q}}+\mathcal{L}[\textbf{q},r]&=0,\nonumber\\
    \dot{\textbf{p}}+\frac{\delta}{\delta\textbf{q}}(\textbf{p}\cdot\mathcal{L}[\textbf{q},r])+\frac{\delta}{\delta\textbf{q}}\mathcal{F}[\textbf{q},r]&=0,\nonumber\\
    \frac{\delta}{\delta r}(\textbf{p}\cdot\mathcal{L}[\textbf{q},r])+\frac{\delta}{\delta r}\mathcal{F}[\textbf{q},r]&=0.
\end{alignat}

The solution to above $\eqref{eq:optimal}$ ($\delta S=0$ and differentiating S w.r.t. p, q and r, respectively) gives the most likely path, denoted by $\bar{\textbf{q}}$, $\bar{\textbf{p}}$, $\bar{r}$, for which $\mathcal{H}$[$\bar{\textbf{q}}$,$\bar{\textbf{p}}$,$\bar{r}$] is a constant of motion. The optimal path can be a local maximum, a local minimum, or a saddle point in the constrained probability space depending on the second variation of action. For the optimal path that represents local maximum, we call it the most likely path or the most probable path.

\section{Quantum Zeno effect in a single qubit-detector system}\label{section 3}

\begin{figure}[htbp!]
    \begin{center}
    \includegraphics[width=0.4\textwidth]{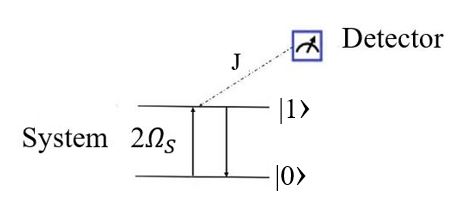}
    \caption{\small{A qubit is performing coherent oscillations between the state $|0\rangle$ and $|1\rangle$ with Rabi frequency, $2\Omega_s$. The state $|1\rangle$ is measured by a detector, another two-level system, with interaction strength $J$, which is initially prepared in $|0\rangle_d$ state and is reset after each measurement.}}
    \label{zeno_system}
    \end{center}
\end{figure}

Let us consider that a qubit is undergoing coherent oscillations between the states $|0\rangle$ and $|1\rangle$ according to the Hamiltonian $H_{(\rm s)}=(1/2)(2\Omega_s)\sigma_{x_{(\rm s)}}$, ($\Omega_s>0$) \cite{minev,parveen}, where, $2\Omega_s$ is the Rabi frequency of the system. As shown in Fig. \ref{zeno_system}, the state $\ket{1}$ of the system is continuously being monitored by the detector (ancilla). The detector is  subjected to measurements at successive intervals of $\delta t$. The detector is another two-level system, initially prepared in the eigenstate, $\ket{0}_{(\rm d)}$ of $\sigma_{z_{\rm (d)}}$ with eigenvalue $1$. We measure $\sigma_{y_{\rm (d)}}$ of the ancilla, after which it is reset to $\ket{0}_{(\rm d)}$. The interaction of the detector with the system is subsumed in a Hamiltonian,
\begin{equation}
    H_{(\rm s-\rm d)}=\frac{J}{2}(\mathbb{I}-\sigma_z)_{(\rm s)}\otimes\sigma_{y_{(\rm d)}},
\end{equation}
where the subscripts $(\rm s)$ and $(\rm d)$ stand for system (qubit) and detector respectively. $J$ is the coupling strength between the detector and the system and the possible measurement outcome, $r$, can be $0$ or $1$. Hence, the system is evolving under the combined effect of its Hamiltonian and the coupling to the detector which is given by the unitary evolution due to the total Hamiltonian, $H=H_{(s)}+H_{(s-d)}$.  
The combined total system-detector Hamiltonian is then,
\begin{equation}
    H=H_{(\rm s)}\otimes{\mathbb{I}}_{(\rm d)}+H_{(\rm s-d)}
\end{equation}
It may be recalled that the Hamiltonian describing Rabi oscillation is time-periodic owing to the time-dependent external field. The evolution is thus describable in terms of a Floquet operator over a period of the drive. This is interrupted by the measurements every $\delta t$ in time. The total evolution operator can be written in a factorized form, owing to the results in \cite{karner,karner1}. For the detector state to be in $|r\rangle_{\rm (d)}$ and in the scaling limit for continuous measurement, $\delta t\to 0$, $J^2\delta t\to\alpha=\rm{constant}$ \cite{parveen}. The state of the system is
\begin{equation}
   \ket{\psi(t+dt)}=M^{(r)}U_{(\rm s)}\ket{\psi(t)}
\end{equation}
where $U_{(\rm s)}=e^{-\iota H_{(\rm s)}\delta t}$ describes the unitary evolution of the system and $M^{(r)}$ is the measurement operator which is given by
\begin{equation}
    M^{(r)}=~_{\rm (d)}\bra{r}e^{-\iota H_{(\rm s-\rm d)}\delta t}\ket{0}_{(\rm d)}.
\end{equation}
Using this, the Kraus operators assume the form \cite{krauss,jordan3},
\begin{align}\label{eq:M}
    M^{(0)}=\begin{bmatrix}
    1 & 0\\
    0 & \cos{J\delta t}
    \end{bmatrix}, \quad
    M^{(1)} =\begin{bmatrix}
    0 & 0\\
    0 & \sin{J\delta t}
    \end{bmatrix}.
\end{align}
For the post-selected dynamics, for $r=0$, we only look at the subensemble evolution,
\begin{equation}
   \ket{\psi(t+\delta t)}=M^{(0)}U_{(\rm s)}\ket{\psi(t)}.
\end{equation}
After evolution, the density matrix of the system will be \cite{jordan,jordan2},
\begin{alignat}{1}\label{rhotplusdt}
    \rho(t+\delta t)&=\frac{M^{(0)}U_{(\rm s)}\ket{\psi(t)}\bra{\psi(t)}|U_{(\rm s)}^\dagger M^{(0)\dagger}}{Tr[M^{(0)}U_{(\rm s)}\ket{\psi(t)}\bra{\psi(t)}|U_{(\rm s)}^\dagger M^{(0)\dagger}]}\nonumber\\
    &=\frac{M^{(0)}U_{(\rm s)}\rho(t)U_{\rm (s)}^\dagger M^{(0)\dagger}}{Tr[M^{(0)}U_{(\rm s)}\rho(t)U_{(\rm s)}^\dagger M^{(0)\dagger}]},
\end{alignat}
where the denominator term is for normalization. For the initial system assuming the general form of density matrix,
\begin{equation}\label{rhot}
    \rho(t)=\frac{1}{2}\begin{bmatrix}
    1+z & x-\iota y\\
    x+\iota y & 1-z
    \end{bmatrix},
\end{equation}
\eqref{rhotplusdt} gives
\begin{equation}\label{eq:rho_tdt}
    \rho(t+\delta t)=\frac{1}{2}\begin{bmatrix}
    1+z+\frac{1}{2}(\alpha(1-z^2)+4\Omega_s y)\delta t & x-\iota y-\frac{1}{2}z((x-\iota y)\alpha-4\iota\Omega_s)\delta t\\
    x+\iota y-\frac{1}{2}z((x+\iota y)\alpha+4\iota\Omega_s)\delta t & 1-z-\frac{1}{2}(\alpha(1-z^2)+4\Omega_s y)\delta t
    \end{bmatrix}.
\end{equation}
Equating both sides for the updated coordinates, we get,
\begin{alignat}{1}\label{eq:update}
    x(t+\delta t)&=x(t)-2\Omega_s\lambda x(t)z(t)\delta t,\\\nonumber
    y(t+\delta t)&=y(t)-2\Omega_s z(t)(1+\lambda y(t))\delta t,\\\nonumber
    z(t+\delta t)&=z(t)+2\Omega_s (\lambda(1-z(t)^2)+y(t))\delta t,
\end{alignat}
where $\lambda=\frac{\alpha}{4\Omega_s}$. If we take the initial $x$-coordinate to be $0$, the update is entirely in $y$-$z$ plane which shows there is no update in $x$-coordinate. This can further be written as
\begin{alignat}{1}
    \dot{x}(t)&=0,\\\nonumber
    \dot{y}(t)&=-2\Omega_s z(t)(1+\lambda y(t)),\\\nonumber
    \dot{z}(t)&=2\Omega_s (\lambda(1-z(t)^2)+y(t)).
\end{alignat}
The equations may be re-written in terms of an angle variable. Writing $y=\sin\theta$ and $z=\cos\theta$, the update equation for $\theta$:
\begin{equation}\label{eq:update_dot}
\dot{\theta}(t)=-2\Omega_s (1+\lambda\sin{\theta(t)}).
\end{equation}
which is the same as found by \cite{parveen}. As explained in Sec. 2, the functional $\mathcal{F}[\textbf{q},r]$ is the coefficient of linear order expansion of the term $Tr[M_0^\dagger M_0\rho]$. This comes out to be 
\begin{equation}
    \mathcal{F}[\textbf{q},r]=-\frac{\alpha}{2}(1-\cos\theta).
\end{equation}
The Hamiltonian can be written according to \eqref{eq:ham},
\begin{alignat}{1}\label{eq:finalham}
    \mathcal{H}=-2\Omega_s[p_\theta(1+\lambda\sin{\theta})+\lambda(1-\cos{\theta})]
\end{alignat}
where $p_\theta$ is the canonical conjugate variable to $\theta$. The corresponding Hamilton's equations are
\begin{alignat}{1}\label{eq:canonical}
    \dot{\theta}(t)&=\frac{\partial \mathcal{H}}{\partial p_\theta}=-2\Omega_s(1+\lambda\sin{\theta}),\nonumber\\
    \dot{p}_\theta(t)&=-\frac{\partial \mathcal{H}}{\partial \theta}=2\Omega_s\lambda( p_\theta\cos\theta+\sin\theta).
\end{alignat}
These equations describe the phase space flow and contain a wealth of information. We now turn to a description and depiction of the flow for different values of $\lambda$. At the outset, it is important to remember that lessons from quantum Zeno effect instruct us of the effect of repeated measurements on the transition probability. In turn, this implies a relation to the ``time taken for transition" - admittedly a term used here to guide our intuition on the basis of statistics rather than to clock the time. We turn to an elaboration of dynamics of the system in regimes separated by values of $\lambda$ about one - guided by the nullclines.  

The connection with Minev's experiment\cite{minev} vis a vis Dehmelt shelving scheme and quantum Zeno effect is brought out in Case II. We shall see that even though in Case I, the duration of the Rabi oscillation period is prolonged, shelving occurs only (Case II) with the appearance of a hyperbolic point (an attractor in $\theta$ on Bloch sphere). The ``third level" in the standard discussion \cite{deh,ludlow} appears as the state corresponding to $\theta_1$. 

\subsection{Case I: \texorpdfstring{$\lambda < 1$}{Lg}}

The transitions would occur at Rabi frequency when the parameter, $\lambda $ is zero. If we interrogate the system and perform a measurement before an oscillation is completed, we would have enhanced the duration of an oscillation. We need to show this, however. It is only in the limit of a very large number of successive measurements, spaced infinitesimally in time, that the transition would be completely arrested, leading to quantum Zeno effect, established by Misra and Sudarshan \cite{ms,peres}. We find it illustrative as well as instructive to draw conclusions by studying the canonical sections in phase space.     

\subsubsection{Portraits of qubit evolution in phase space}

\begin{figure*}[htbp!]
    \centering
    \subfloat[]{\includegraphics[width=0.42\textwidth]{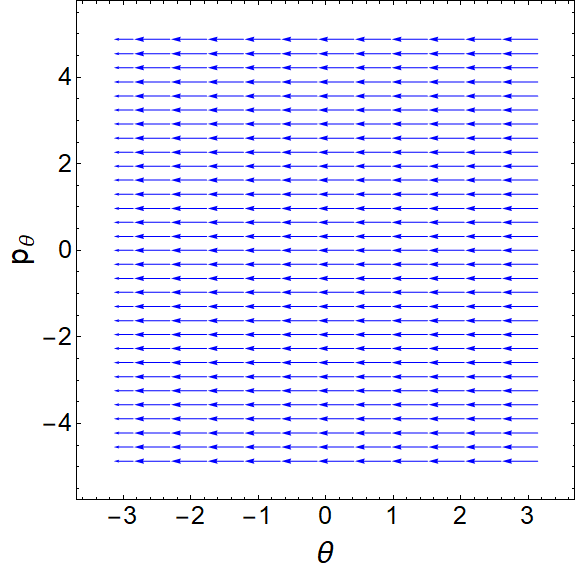}}\hspace{5mm}
    \subfloat[]{\includegraphics[width=0.42\textwidth]{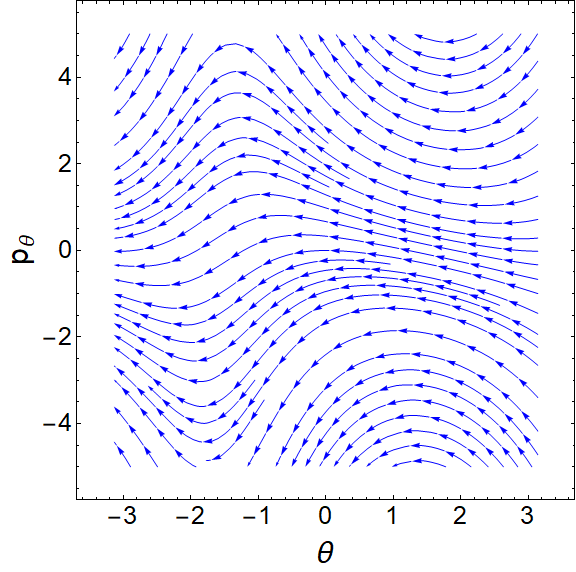}}\\
    \caption{Phase space plots after extremizing the action showing invariant curves (most optimal paths) where the Hamiltonian, ${\mathcal H}(\theta, p_{\theta})$ = constant \eqref{eq:finalham}. (a) For  $\lambda=0$, we notice that $p_{\theta}$ is a constant of the motion; (b) for $\lambda=0.5$, while energy is constant, $p_{\theta}$ is no longer a constant ($\Omega_s=0.5$). It is evident that $\theta$ is continuously evolving towards $-\pi$, when $\lambda$ is between 0 and 1.} 
    \label{fig:Phase_space<1}
\end{figure*}

We consider curves in $({\theta, p_{\theta}})$-phase plane corresponding to the evolution of the qubit from $\theta=0$ (state $|0\rangle$) to $\theta=-\pi$ (state $|1\rangle$) on the Bloch sphere. For $\lambda $ equal to zero, Fig. \ref{fig:Phase_space<1}(a) shows that $p_{\theta}$ is a constant of the motion and frequency is $2\Omega_s$, the Rabi frequency. For a non-zero $\lambda$, the straight lines in Fig. \ref{fig:Phase_space<1}(b) become unstable in a way that there is a pair of $p_{\theta}$-values about which there is attraction (repulsion) for positive (negative) $\theta$. 

From \eqref{eq:finalham}, for $-\mathcal{H}/(2\Omega_s) = {\mathcal E}$, we have
\begin{equation}\label{eq:ptheta}
p_{\theta}(\theta ; \lambda , \mE) = \frac{\mE - \lambda (1 - \cos \theta)}{1 + \lambda \sin \theta}.     
\end{equation}

\subsubsection{Action integral}

We calculate the action for the phase space curves traced by the dynamics dictated by the Hamiltonian, \eqref{eq:finalham}. The stochastic action integral for the system is given by
\begin{alignat}{1}\label{eq:area_la<1}
\mA (\lambda ) &= \int_{t_i}^{t_f}(-p_\theta \dot{\theta}+\mathcal{H})dt\\
&=\int_{\theta_i}^{\theta_f}\mathcal{F}\frac{dt}{d\theta}d\theta\\
&= \frac{2\lambda}{\sqrt{1-\lambda^2}}\bigg(\tan^{-1}\bigg[\frac{\lambda+\tan{\frac{\theta_f}{2}}}{\sqrt{1-\lambda^2}}\bigg]-\tan^{-1}\bigg[\frac{\lambda+\tan{\frac{\theta_i}{2}}}{\sqrt{1-\lambda^2}}\bigg]\bigg)-\ln\bigg[\frac{1+\lambda\sin{\theta_f}}{1+\lambda\sin{\theta_i}}\bigg].
\end{alignat}
This is seen to change sign about $\lambda = {\mathcal E}$. Fig. \ref{fig:Phase_space<1} (b) shows the curves with negative (positive) areas in the part where $p_{\theta}$ is positive (negative).

\subsubsection{Time of transition}\label{time 0f transition}

As discussed earlier, transition probability is intimately related to the time of transition from $\theta=0$ to $\theta=-\pi$ which becomes interesting in the context of any discussion of Zeno effect. In this case, it is simply calculated using \eqref{eq:canonical}:
\begin{equation}\label{eq:timeperiod_la<1}
    T_{\lambda<1}=\frac{\frac{\pi}{1-\lambda^2}+\frac{2\tan^{-1}\left[\frac{\lambda}{\sqrt{1-\lambda^2}}\right]}{\sqrt{1-\lambda^2}}}{2\Omega_s}.
\end{equation}
For $\lambda=0$, this comes out to be $T_{\lambda<1}=\frac{1}{2}(\frac{2\pi}{2\Omega_s})$ which, as expected, turns out to be exactly half the time-period of the Rabi oscillations. This is independent of the energy, $\mE$, indicating that on all constant energy curves in the phase space, this time of transition remains constant. Although it might seem surprising, but it is easily understood by the fact that the action as well as momentum are linear in ${\mathcal E}$.  

\begin{figure}[htbp!]
    \begin{center}
    \includegraphics[width=0.5\textwidth]{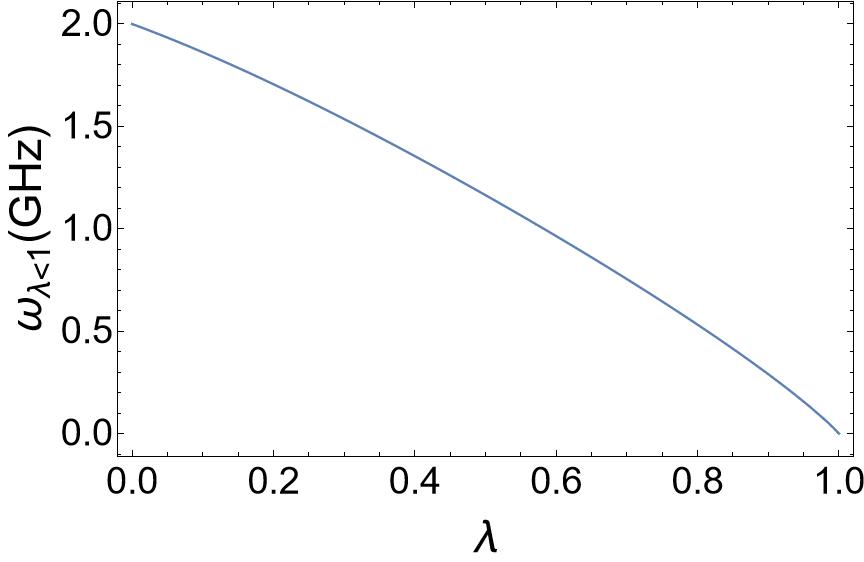}
    \caption{Plot shows the frequency of transitions (in the unit of GHz) from $\theta=0$ to $\theta=-\pi$ with $\lambda$ for $\lambda<1$. The frequency decreases with increasing $\lambda$, indicating that system makes fewer oscillations on increasing the detection frequency and comes to rest at $\lambda = 1$, marking the onset of Zeno regime.}
    \label{time_la<1}
    \end{center}
\end{figure}

The time of making a transition from state $|0\rangle$ to $|1\rangle$ is longer than the Rabi oscillations for $\lambda=0$. Fig. \ref{time_la<1} brings out the variation in transition frequency as detection frequency increases. It takes longer and longer for the oscillations to complete (Fig. 9(a)). Eventually, at $\lambda = 1$, a kind of resonance condition is satisfied when Rabi period is twice the time between successive detections; here, frequency of oscillations vanishes.  The system remains in the initial state, the quantum Zeno effect sets in. Similar conclusion is drawn in \cite{parveen}. 

\subsection{Case II: \texorpdfstring{$\lambda > 1$}{Lg}}

This regime is clearly Zeno regime where a cascade of stages have been shown to occur \cite{parveen}. Inspired by this and the beautiful and exciting experiment \cite{minev} on the possibility of controlling quantum jumps, we present our exploration of this phenomenon in phase space, as explained above. In this description, via action integral, it is possible to make estimates on transition times. The cascades also show up when the description appears in phase space. As above, we begin the description with critical points about which the phase space curves would be organized.  

\subsubsection{Critical points and their stability}\label{Critical points}

Critical points or the nullclines are obtained by setting the right hand side of \eqref{eq:update_dot}  to zero. We would like to show the points of equilibrium for this system on the Bloch sphere as well. Critical points $\theta_1$ and $\theta_2$ for $\lambda > 1$ turn out to be
\begin{alignat}{1}\label{eq:critical_th}
  \theta_1=-\sin^{-1}\frac{1}{\lambda}, \quad  \theta_2=\sin^{-1}\frac{1}{\lambda}-\pi.
\end{alignat}
From \eqref{eq:canonical}, $\dot{p}_\theta=0$ gives the critical points in  $p_\theta$:
\begin{alignat}{1}\label{eq:critical_p}
     {p_\theta}_1=\frac{1}{\sqrt{\lambda^2-1}}, \quad {p_\theta}_2=-\frac{1}{\sqrt{\lambda^2-1}}.
\end{alignat}

\begin{figure}[htbp!]
    \begin{center}
    \includegraphics[width=0.5\textwidth]{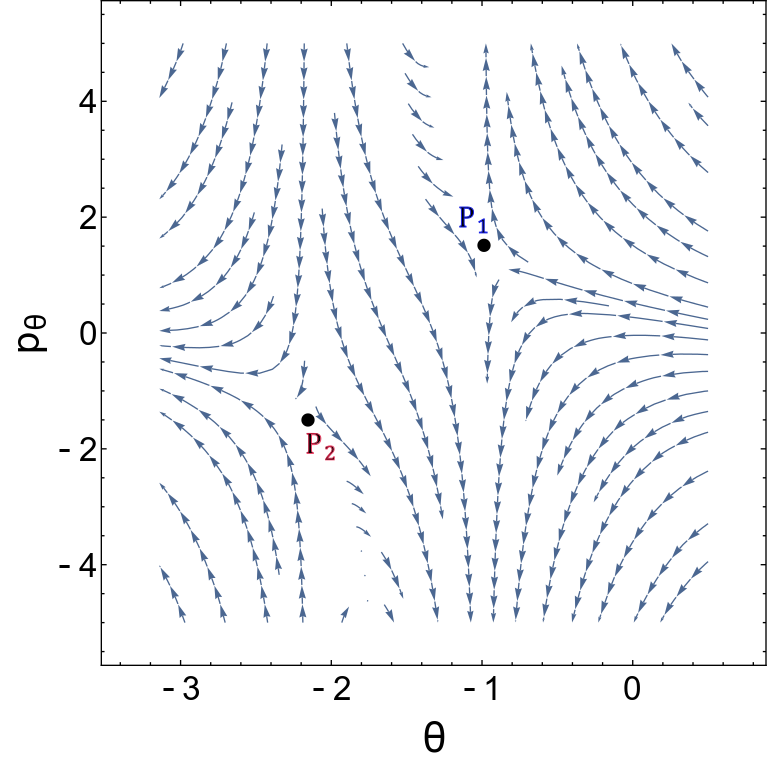}
    \caption{Phase space vector plots for $\lambda=1.2$ is shown for energy, $\mE=2$ with $\Omega_s=0.5$. The plot displays the critical points at $\theta_1=-\sin^{-1}\frac{1}{\lambda}$ and $\theta_2=\sin^{-1}\frac{1}{\lambda}-\pi$, so that $P_1\equiv(-0.985,1.507)$ and $P_2\equiv(-2.156,-1.507)$. At $\theta_1$ ($\theta_2$), the system is stable (unstable), whereas at $p_{\theta_1}$ ($p_{\theta_2}$), system is unstable (stable). Thus, the phase portrait has two saddle points.} 
    \label{fig:Phase_space_>1}
    \end{center}
\end{figure}

To ascertain the stability of these points, we linearize the equations by introducing a small perturbation, $\epsilon_i$ and retain the terms in the resulting equation up to linear order.  Thus, writing   $\theta_1=-\sin^{-1}\frac{1}{\lambda} + \epsilon_1$ and $\theta_2=\sin^{-1}\frac{1}{\lambda}-\pi + \epsilon_2$, we obtain for $\epsilon_i$:
\begin{alignat}{1}\label{eq:stable_th}
    \epsilon_1&=\exp(-2\Omega_s t\sqrt{\lambda^2-1}), \quad \epsilon_2=\exp(2\Omega_s t\sqrt{\lambda^2-1}).
\end{alignat}
which shows that $\theta_1$ is a stable point while $\theta_2$ is an unstable point. 
A similar analysis can be carried out for $p_\theta$. Writing  ${p_\theta}_1=\frac{1}{\sqrt{\lambda^2-1}} + \delta_1$ and ${p_\theta}_2=-\frac{1}{\sqrt{\lambda^2-1}} + \delta_2$, we obtain
\begin{alignat}{1}\label{eq:stable_p}
    \delta_1&=\exp(2\Omega_s t\sqrt{\lambda^2-1}), \quad \delta_2=\exp(-2\Omega_s t\sqrt{\lambda^2-1}),
\end{alignat}
which remarks that the dynamics is unstable about ${p_{\theta}}_1$ and stable about ${p_{\theta}}_2$, which is opposite in sense to that of fixed points of $\theta$. Hence the points determined by $(\theta_1,p_{\theta_1})$ and $(\theta_2,p_{\theta_2})$ are saddle points.

Phase space trajectory of the system is plotted in Fig \ref{fig:Phase_space_>1}, which shows that the system is evolving from $\theta=0$ (state $|0\rangle$) to $\theta_1=-\sin ^ {-1} \frac{1}{\lambda}$ instead of making transitions from $|0\rangle$ to $|1\rangle$. It is staying at $\theta_1$ which is like a metastable state whose lifetime is longer. On increasing the value of interaction parameter (or the detection frequency) as $\lambda\to\infty$, $p_\theta\to\tan\frac{\theta}{2}$. At $p_{\theta_{\lambda\to\infty}}$ the system is freezing in state $|0\rangle$ manifesting the Zeno effect which is desirable for quantum error correction and manipulation of qubits.

On differentiating \eqref{eq:ptheta} w.r.t. $\lambda$ about the unstable point, we obtain a relation between $\lambda$, thereby giving two critical values for $\mE$,
\begin{align}\label{lambdavsE}
    \mE&=\lambda\pm\sqrt{\lambda^2-1}
\end{align}

\begin{figure}[htbp!]
    \begin{center}
    \includegraphics[width=0.7\textwidth]{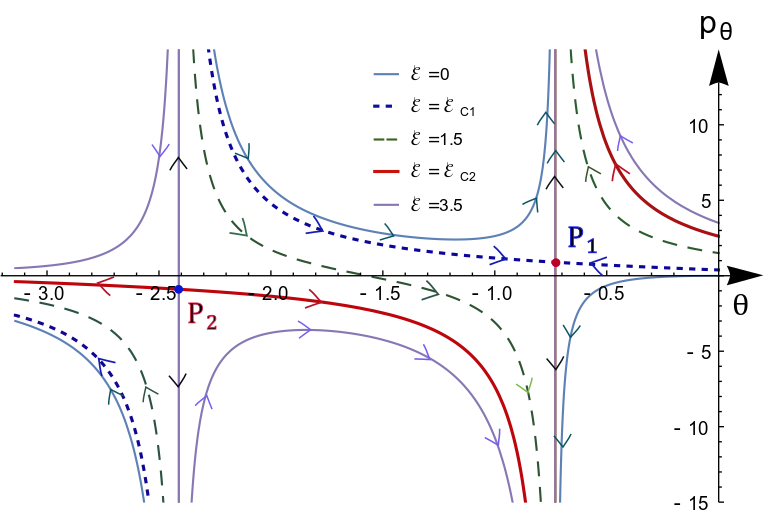}
    \caption{Phase space plot obtained after extremization of stochastic action shows most optimal paths for Zeno regime ($\lambda>1$). From \eqref{lambdavsE}, for $\lambda=1.5$, the plot is for energy values in three ranges, $\mE<\mE_{C1}$, $\mE_{C1}\leq\mE\leq\mE_{C2}$ and $\mE>\mE_{C2}$ ($\Omega_s=0.5$), where $\mE_{C1}$ and $\mE_{C2}$ are the critical values of $\mE$ corresponding to a fixed $\lambda$ ($>1$). We have two separatrices, which cannot be crossed, regardless of the energy. The points $P_1\equiv(-0.729,0.894)$ and $P_2\equiv(-2.411,-0.894)$ are the critical points corresponding to $\lambda=1.5$. The small dashed (bold) line corresponds to energy value $\mathcal{E}_{C1}$ ($\mathcal{E}_{C2}$) and the line with bigger dashes corresponds to the energy value between these two ($\mathcal{E}_{C2}=1.5$). If the system is prepared in state $\ket0$ or in equal superposition of $\ket0$ and $\ket1$, it will evolve to $\theta_1$ and will be localized at that point.}
    \label{fig:separatrix}
    \end{center}
\end{figure}

At these points, we obtain two separatrices from the two critical values of $\mE$. This can be seen in Fig. \ref{fig:separatrix}. Clearly, such situations are only possible for $\mE>0$ as $\lambda$ is a positive quantity. As seen in Fig. 5, the two critical points are such that one is a saddle point with unstable (manifold) direction along $p_{\theta}$ and a stable transverse (manifold) direction almost along $\theta$; the other one is a saddle point with opposite directions. Localization or stabilization of qubit in the neighbourhood of one saddle and destabilization of qubit in the other neighbourhood drives the dynamics - a projection of that is seen on the Bloch sphere. Fig. 5 shows seven regions in which the phase flow is divided and organized thus due to saddles and separatrices. These curves portray the most likely behaviour of states. A detailed description of the possible paths may be found in a rather insightful work by Chantasri and Jordan \cite{jordan2} where they have also studied qubit stabilization in the case of qubit measurement with linear feedback. In their case, the critical point are stable in $\theta$. In our case, in quantum Zeno regime, stabilization around $\theta_1$ and destabilization around $\theta_2$ is clearly seen, these are special states other than $|0\rangle$ and $|1\rangle$. However, the quantum fluctuations and tunneling across the separatrices which would enable the system to continue evolving. The tunneling probability is proportional to $\exp [- \tau S_I/\hbar]$ where $S_I$ is the imaginary part of the action as the separatrix is crossed and $\tau $ is the inverse of the characteristic stability exponent along the unstable direction \cite{alonso,ssj} found by the linearization process above.         

\begin{figure*}[htbp!]
    \centering
    \subfloat[]{\includegraphics[width=0.45\textwidth]{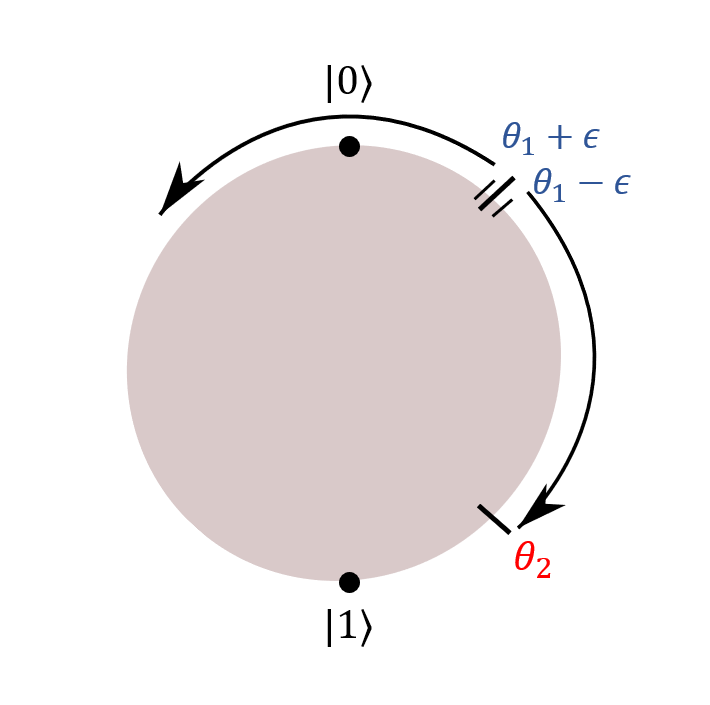}}
    \subfloat[]{\includegraphics[width=0.45\textwidth]{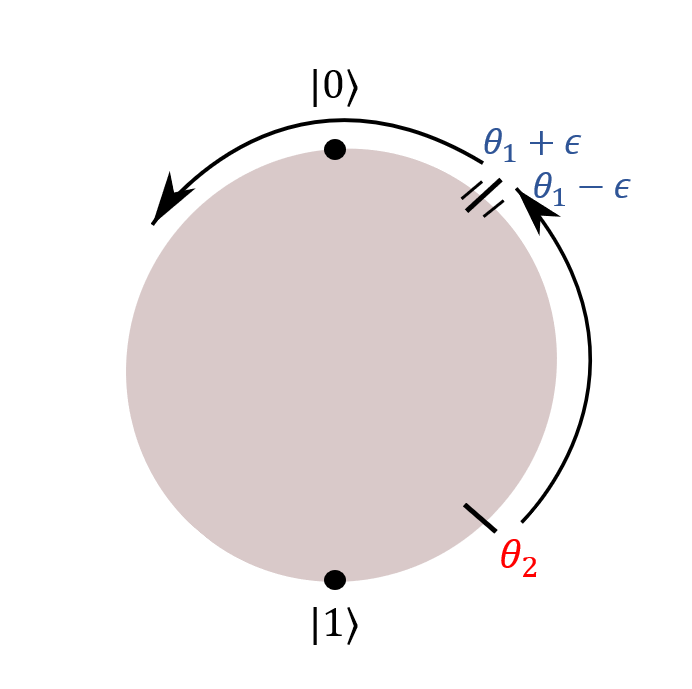}}
    \caption{To see the discontinuity in action, we evaluate action as a function of an arbitrary angle, $\epsilon$, and $\mu=\lambda>1$ in the region $\theta_i=\theta_1+\epsilon\to\theta_f=\theta_2$ and $\theta_i=\theta_1-\epsilon\to\theta_f=\theta_2$ in Fig. (a) and in the region $\theta_i=\theta_2\to\theta_f=\theta_1-\epsilon$ and $\theta_i=\theta_1+\epsilon\to\theta_f=\theta_2$ in Fig. (b) for $\lambda=1.5$ and $\epsilon\to0$.} 
    \label{fig:blochsphere}
\end{figure*}

\begin{figure*}[htbp!]
    \centering
    \subfloat[]{\includegraphics[width=0.48\textwidth]{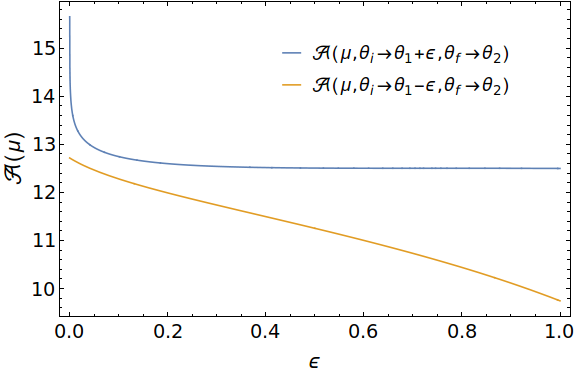}}\hspace{2mm}
    \subfloat[]{\includegraphics[width=0.5\textwidth]{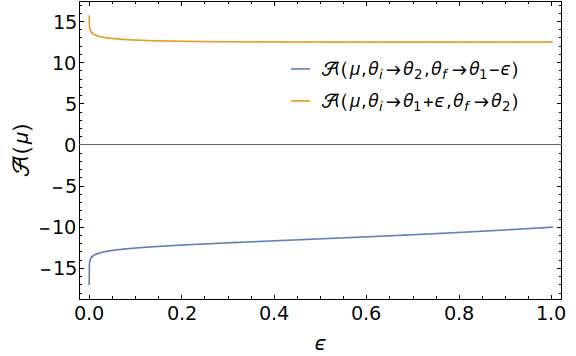}}
    \caption{Fig. (a) represents, for an arbitrary constant $\epsilon\to0$ and $\lambda=1.5$, action evaluated from $\theta_i\to\theta_1+\epsilon$ to $\theta_f\to\theta_2$ (blue) and from $\theta_i\to\theta_1-\epsilon$ to $\theta_f\to\theta_2$ (orange) for the paths shown in Fig. \ref{fig:blochsphere}(a) and (b) respectively. These paths comply with the direction of stability near the points $\theta_1$ and $\theta_2$ as in the vector plots shown in Fig. \ref{fig:Phase_space_>1} and hence have the opposite direction but same sign. Whereas, in Fig. (b), action for both the paths is calculated in the same direction, which is opposite in the sense of the direction of stability on one side of the stable point, and hence have opposite signs.} 
    \label{Area}
\end{figure*}

\subsubsection{Action and quantum jumps}

We show here that associated with a quantum jump, there appears a discontinuity in action. 
For $\lambda>1$, let us denote it by $\mu$, $\sqrt{1 - \lambda^2} = i\sqrt{\mu^2 - 1}$. Employing the identity \cite{ablowitz},
\begin{equation}
    \tan^{-1} \xi = \frac{1}{2i}\log \frac{i - \xi}{i + \xi}, 
\end{equation}
we have action upon integration from an initial point $\theta_i$ to $\theta_f$:
\begin{alignat}{1}\label{eq:area_la>1}
\mA (\mu ) &=-\frac{\mu}{\sqrt{\mu^2-1}}\log\left[\bigg(\frac{\sqrt{\mu^2-1}+\mu+\tan\frac{\theta_f}{2}}{\sqrt{\mu^2-1}-\mu-\tan\frac{\theta_f}{2}}\bigg)\bigg(\frac{\sqrt{\mu^2-1}-\mu-\tan\frac{\theta_i}{2}}{\sqrt{\mu^2-1}+\mu+\tan\frac{\theta_i}{2}}\bigg)\right]-\log\left[\frac{1+\mu\sin\theta_f}{1+\mu\sin\theta_i}\right]
\end{alignat}
for $\mu = \lambda > 1$.

Consider the Bloch sphere diagrams in Fig. \ref{fig:blochsphere} (a) and (b). If we plot the area w.r.t. to an arbitrary angle, $\epsilon\to0$ from a point where initial Bloch angle is $\theta_i=\theta_1-\epsilon$ to $\theta_f=\theta_2$ (clockwise) and the action from $\theta_i=\theta_1+\epsilon$ to $\theta_f=\theta_2$ (anti-clockwise), both are in the opposite direction, but we find that both the action have positive values. If, instead we plot the action in the same direction (anti-clockwise) from $\theta_i=\theta_2$  to $\theta_f=\theta_1-\epsilon$ and $\theta_i=\theta_1+\epsilon$ to $\theta_f=\theta_2$, we find that they have opposite signs as shown in Fig. \ref{Area}. This displays a discontinuity in the area at the stable point $\theta_1$, thus implying a discontinuity in action at that point. This discontinuity in action may be interpreted as a quantum jump. 

\begin{figure}[htbp!]
    \begin{center}
    \includegraphics[width=0.7\textwidth]{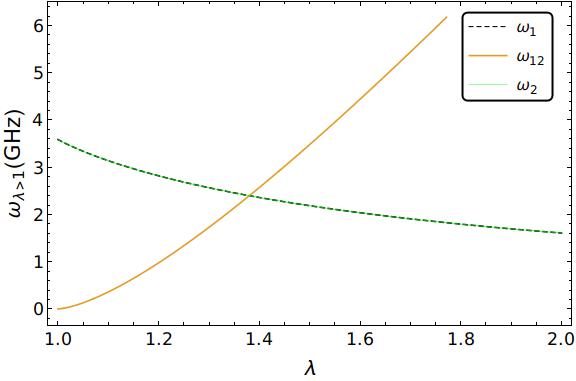}
    \caption{Plot shows the frequency of transitions from $\theta=0$ to $\theta=-\pi$ with $\lambda$. Here, $\omega_1$ (dashed black) is the frequency of transitions from $\theta=0\to\theta_1$, $\omega_{12}$ (orange) is the frequency of transitions from $\theta=\theta_2\to\theta_1$ and $\omega_2$ (light green) is the frequency of transitions from $\theta=\theta_2\to-\pi$. Frequencies, $\omega_1$ and $\omega_2$ decrease at the same rate with increasing $\lambda$ (hence the plots of these frequencies are overlapped), indicating that system makes fewer oscillations on increasing the detection frequency. However, $\omega_{12}$ increases with increasing $\lambda$ which shows that system transits at a faster rate from $\theta_2\to\theta_1$.}
    \label{time_la>1}
    \end{center}
\end{figure}

\begin{figure*}[htbp!]
    \centering
    \subfloat[]{\includegraphics[width=0.46\textwidth]{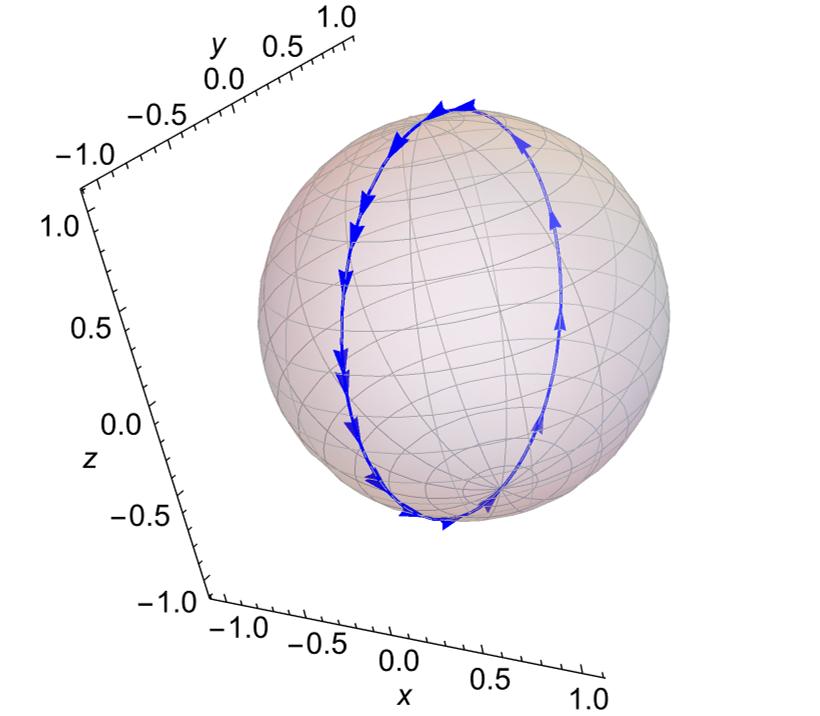}}
    \subfloat[]{\includegraphics[width=0.43\textwidth]{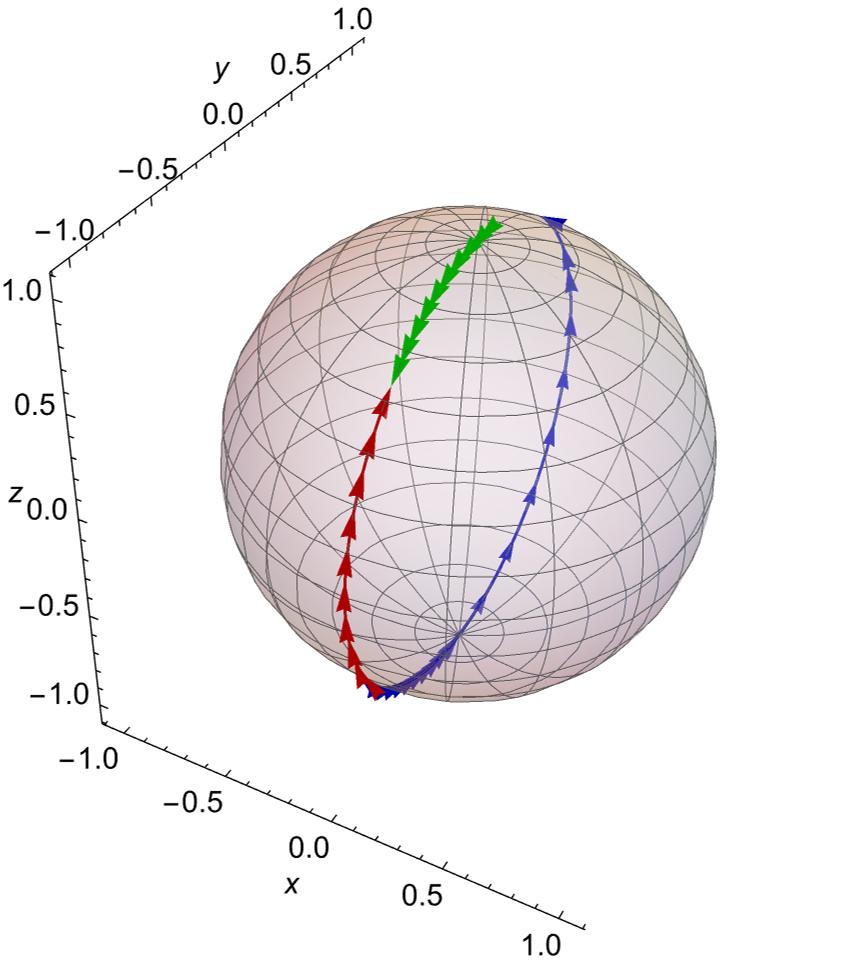}}\\
    \caption{(a) For $\lambda = 0.423 (< 1)$, prolonged Rabi oscillations are shown here in accordance with \eqref{eq:timeperiod_la<1}; (b) For $\lambda = 1.5 (> 1)$, in the Zeno regime, the trajectory is divided into three distinct arcs due to the presence of stable and unstable points. } 
    \label{sphere_projective}
\end{figure*}

\begin{figure}[htbp!]
    \begin{center}
    {\includegraphics[width=0.6\textwidth]{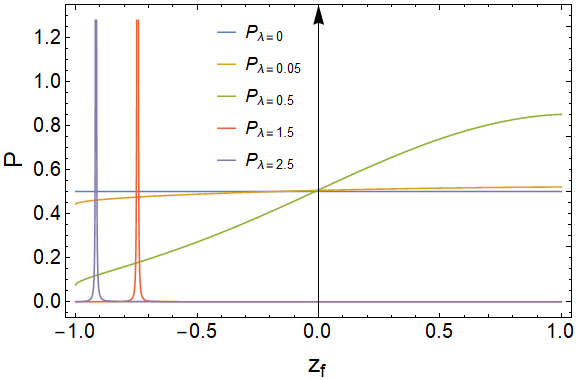}}
   \caption{The extremization of action for both the conditions of $\lambda$ indicates that for $\lambda=0$, the probability density $P(z_f|z_I)$ remains constant. For very small values of $\lambda$, i.e., low detector frequencies ($\lambda=0.05, 0.5$), the qubit traverses the path from initial state $z_i=1$ ($\theta_i=0$) to final state $z_f=-1$ ($\theta_f=-\pi$). However, when $\lambda$ is increased beyond $1$, ($\lambda=1.5, 2.5$), the qubit starts from the initial state and arrested at $z_1$, corresponding to the critical point of \eqref{eq:critical_th}, which is a function of $\lambda$. }
    \label{fig:expS}
    \end{center}
\end{figure}

We plot the probability density, i.e., the exponential of extremized action \eqref{eq:area_la<1}, \eqref{eq:area_la>1} with respect to $z_f$ (Fig. \ref{fig:expS}) to see the leading term of $P(z_f|z_I)$, starting from $\theta_i$=$0$ or $z_i=1$. For large value of $\lambda$=$2.5$, final state is mostly around the stable states. As the value of $\lambda$ decreases the curve gets broader and the most probable final states move toward $z_f$=$-1$.

Plotted probability densities are normalized as the area under the curve is $1$ for all the curves corresponding to different values of $\lambda$. For $\lambda=0$, the probability is constant for all the values of $z_f$ which shows the uninterrupted Rabi oscillations of the qubit. As the value of $\lambda$ increases, the probability of reaching the final state $z_f=-1$ (or $\theta=\pi$) decreases, which corresponds to the prolonged Rabi oscillations of the qubit. For $\lambda > 1$, the probability density shows almost $0$ value for all the $z_f$ except at the ``critical points" of \eqref{eq:critical_th}, where it has peaks. This shows that the qubit is arrested at these stable points exhibiting the quantum Zeno effect.

\subsubsection{Time of transition}

For $\lambda>1$, the  time of transition cannot be calculated over the range of $\theta$ as there are points where the phase space curves exhibit discontinuities. However, we can divide the interval into three parts:  $0\to\theta_1+\epsilon$, $\theta_2+\epsilon\to\theta_1-\epsilon$ and $\theta_2-\epsilon\to-\pi$,corresponding to frequencies, $\omega_1$, $\omega_{12}$ and $\omega_2$, respectively, where $\epsilon$ is an arbitrary small number. Frequencies $\omega_1$ and $\omega_2$ decrease with increasing $\lambda$, indicating that system is making fewer oscillations on increasing the detection frequency and comes to rest at $\lambda\gg1$, marking the end of Zeno regime as the system freezes completely. Whereas, $\omega_{12}$ increases with increasing $\lambda$ which shows that if the system is anywhere between $\theta_1$ and $\theta_2$ it transits at a faster rate from $\theta_2\to\theta_1$, see Figs. \ref{time_la>1}, \ref{sphere_projective}(b).

\section{Diffusive measurement}\label{Diffusive}

We consider the system discussed in Sec. \ref{section 3}, albeit with a measurement model . The state $\ket{1}$ of the qubit is being continuously monitored by a detector (ancilla) which has a characteristic time $\tau$ and the strength of interaction between the qubit and the detector is $J=\sqrt{\frac{\alpha}{\delta t}}$, where $\alpha$ is a constant \cite{parveen}. The detector is another two-level system, initially prepared in $\ket{0}_{(\rm d)}$ state of $\sigma_{{z}_{(\rm d)}}$ and is reset prior to each measurement of the observable $\sigma_{y_{(\rm d)}}$. For a total time of measurement, $T=n \delta t$, divided into $n$ intervals such that measurement readout is obtained at each interval $\delta t$, weak coupling implies, $\tau \gg T$ \cite{jordan2} and $\tau \gg\delta t$ \cite{ashhab}, so that the detector's operation during each interval is independent. 

The total system-detector Hamiltonian is
\begin{equation}\label{eq:ham_diffusive}
    H=\Omega_s\sigma_{{x}_{(\rm s)}}\otimes\mathbb{I}_{(\rm d)}+\frac{J}{2}(\mathbb{I}-\sigma_z)_{(\rm s)}\otimes\sigma_{y_{(\rm d)}}.
\end{equation}

The combined system evolves via the evolution operator $\mathcal{U}$, which can be decomposed into operators corresponding to Schr\"{o}dinger evolution of the system, interaction between system and detector, followed by measurement. The  evolution of density matrix can be written as: \begin{equation}\label{eq:update_diff}
    \rho(t+dt)=\frac{\mathcal{U}\rho(t)\mathcal{U}^\dagger}{Tr[\mathcal{U}\rho(t)\mathcal{U}^\dagger]}.
\end{equation}
Let us consider the free evolution of a closed system which is described by the time-independent Schr\"{o}dinger equation
\begin{equation}
    \rho(t+dt)=U\rho(t)U^\dagger,
\end{equation}
where $U(t) = e^{-\iota H dt}$. The weak measurement of the system, i.e. projective measurement of the ancilla changes the state from $\rho(t)\otimes\ket{\phi}\bra{\phi}$ to $\rho(t+dt,i)\otimes\ket{i}\bra{i}$ \cite{martin}
\begin{equation}
    \rho(t+dt,i)\otimes\ket{i}\bra{i}=\frac{({\mathcal I}_{(\rm s)}\otimes \ket{i}\bra{i})U(\rho(t)\otimes\ket{\phi}\bra{\phi})U^\dagger({\mathcal I}_{(\rm s)}\otimes\ket{i}\bra{i})}{Tr[({\mathcal I}_{(\rm s)}\otimes \ket{i}\bra{i})U(\rho(t)\otimes\ket{\phi}\bra{\phi})U^\dagger({\mathcal I}_{(\rm s)}\otimes\ket{i}\bra{i})]},
\end{equation}
where $\ket{i}\bra{i}$ is the projection operator into the $ith$ eigenspace of the observable subjected to measurement, and $\ket{\phi}$ is the initial state of the ancilla. 

We perform a projective measurement on the ancilla and trace over the ancilla to obtain the final state of the system:
\begin{alignat}{1}\label{eq:rho_martin}
\rho(t+dt,i)&=\frac{({\mathcal I}_{(\rm s)}\otimes \langle i|)U(\rho(t)\otimes|\phi\rangle\langle\phi|)U^\dagger({\mathcal I}_{(\rm s )}\otimes|i\rangle)}{Tr[({\mathcal I}_{(\rm s)}\otimes \langle i|)U(\rho(t)\otimes|\phi\rangle\langle\phi|)U^\dagger({\mathcal I}_{(\rm s)}\otimes|i\rangle)]}\nonumber\\
&=\frac{{\mathcal M}_i\rho(t){\mathcal M}_i^\dagger}{Tr[{\mathcal M}_i\rho(t){\mathcal M}_i^\dagger]},
\end{alignat}
where ${\mathcal M}_i$'s are the Kraus operators. ${\mathcal M}_i=|i\rangle\langle i|$ is projection operator with $\sum_i{\mathcal M}_i=\hat{I}$. In \eqref{eq:rho_martin}, ${\mathcal M}_i=({\mathcal I}_{s}\otimes\langle i|)U({\mathcal I}_{s}\otimes |\phi\rangle)$ is the Kraus operator acting on the system, given by
\begin{alignat}{1}
{\mathcal M}_i & = ({\mathcal I}_{(\rm s)}\otimes\bra{i}) U({\mathcal I}_{(\rm s)}\otimes\ket{g})\nonumber\\
&=({\mathcal I}_{(\rm s)}\otimes\bra{i}|) \exp{\left[-\iota \Omega_s\sigma_{{x}_{(\rm s)}}\otimes\mathbb{I}_{(\rm d)}\delta t-\iota J\bigg(\frac{{\mathcal I}_{(\rm s)}-\sigma_{{z}_{(\rm s)}}}{2}\otimes\sigma_{{y}_{(\rm d)}}\bigg)\delta t\right]}({\mathcal I}_{(\rm s)}\otimes\ket{g})\nonumber\\
&=({\mathcal I}_{(\rm s)}\otimes\langle i|)\bigg[{\mathcal I}_{(\rm s)}\otimes {\mathcal I}_{(\rm d)}-\iota\Omega_s\sigma_{{x}_{(\rm s)}}\otimes\mathbb{I}_{(\rm d)}\delta t-\iota J\delta t \frac{{\mathcal I}_{(\rm s)}-\sigma_{{z}_{(\rm s)}}}{2}\otimes\sigma_{{y}_{(\rm d)}}\nonumber\\
&-(J\delta t)^2\frac{{\mathcal I}_{(\rm s)}-\sigma_{{z}_{(\rm s)}}}{4}\otimes {\mathcal I}_{(\rm d)}\bigg]({\mathcal I}_{(\rm s)}\otimes|g\rangle)+\mathcal{O}(\delta t^3).
\end{alignat}
Projecting the ancilla in the eigenstates of $\sigma_y$, i.e., $|i\rangle=|\pm y\rangle$, leads to the following form of the Kraus operators:
\begin{alignat}{1}
{\mathcal M}_{\pm}& =\frac{1}{\sqrt{2}}\bigg({\mathcal I}_{(\rm s)} \mp \iota J\delta t \frac{{\mathcal I}_{(\rm s)}-\sigma_{{z}_{(\rm s)}}}{2}-\iota \Omega_s\sigma_{{x}_{(\rm s)}}\delta t- \frac{(J\delta t)^2}{2}\frac{{\mathcal I}_{(\rm s)}-\sigma_{{z}_{(\rm s)}}}{4}\bigg)\nonumber\\
&=\frac{1}{\sqrt{2}}\bigg({\mathcal I}_{(\rm s)} \mp \iota\sqrt{\alpha \delta t} \frac{{\mathcal I}_{(\rm s)}-\sigma_{{z}_{(\rm s)}}}{2}-\iota\Omega_s \sigma_{{x}_{(\rm s)}}\delta t- \alpha \delta t\frac{{\mathcal I}_{(\rm s)}-\sigma_{{z}_{(\rm s)}}}{4}\bigg).
\end{alignat}

Due to inherent stochasticity in the detector, the randomness of measurement outcomes finds description in stochastic calculus. Defining these by a random variable, $W(t)$ such that 
\begin{alignat}{1}
W(t+\delta t) = W(t)\pm\sqrt{\delta t}
\end{alignat}
with $W(0)=0$. In the continuum limit, $\delta t\to 0$, $W(t)$ describes a Wiener process. Wiener increment, $dW$ is a zero-mean Gaussian distributed random variable with variance $\delta t$. We can write the updated state (after dropping out the subscript (\rm s) as the Kraus operator is operating on the system only) as \cite{martin}:
\begin{alignat}{1}\label{update1}
    \ket{\bar{\psi}(t+\delta t)}&=\sqrt{P(dW)}{\mathcal M}_{dW}|\psi(t)\rangle\nonumber\\
    &=\sqrt{P(dW)}\bigg({\mathcal I} -\iota \sqrt{\alpha}dW \frac{({\mathcal I}-\sigma_z)}{2}-\iota\Omega_s \sigma_x \delta t- \alpha \delta t\frac{({\mathcal I}-\sigma_{z})}{4}\bigg)\ket{\psi(t)}
\end{alignat}
This is a linearised stochastic Schr\"{o}dinger equation, where $|\bar{\psi}(t+\delta t)\rangle$ is an unnormalized state. To write the equation with correct statistics, we incorporate an \^{I}to random variable with the same statistics as that of an actual measurement. The probabilities for each of the two possible outcomes corresponding to the eigenstates of $\sigma_y$ are:
\begin{alignat}{1}\label{eq:probability}
P(\pm)&=\langle\psi(t)|{\mathcal M}_{\pm}^\dagger {\mathcal M}_{\pm}\ket{\psi(t)} = \frac{1}{2}.
\end{alignat}
For writing the equation of motion, we replace the random variable $dW^2$=$\delta t$ with an \^{I}to random variable $\delta r$. Averaging with the probability distribution \eqref{eq:probability} gives:
\begin{alignat}{1}
\langle \delta r \rangle &= \sqrt{\delta t}[P(+)-P(-)]= 0\\
{\rm var}(\delta r)& = \langle (\Delta r)^2\rangle-(\langle \Delta r\rangle)^2=\delta t +{\mathcal O}(\delta t)^2
\end{alignat}
Suppose we sum the values of $\delta r$ over many time steps, keeping $\delta t$ small enough so that $\langle \delta r \rangle$ remains constant over a large number of time-steps. By the Central limit theorem, the sum follows a Gaussian distribution with zero mean and variance, $\delta t$. In the limit $\delta t\to 0$, we can write from \cite{lewalle}:
\begin{equation}
    r=\frac{dW}{dt}\sqrt{\tau}.
\end{equation}
The Kraus operator is 
\begin{equation}
    {\mathcal M}_{dW}= \sqrt{P(dW)} \exp \bigg(-\iota\sqrt{\alpha}\frac{r dt}{\sqrt{\tau}} \frac{({\mathcal I}-\sigma_{z})}{2}-\iota\Omega_s \sigma_x dt-\alpha\frac{({\mathcal I}-\sigma_{z})}{4} dt\bigg).
\end{equation}
For a  large number of steps, $N \to \infty$, the prefactor $\sqrt{P(dW)}$ limits to a Gaussian which leads to the following Kraus operator, 
\begin{alignat}{1}
    {\mathcal M}_{r}&= \bigg(\frac{dt \, e^{-r^2 dt/\tau}}{2\pi \sqrt{\tau}}\bigg)^{1/4} \exp \bigg(-\iota\Omega_s\sigma_x dt-\iota \sqrt{\frac{\alpha}{\tau}}\frac{({\mathcal I}-\sigma_{z})}{2} r dt-\alpha\frac{({\mathcal I}-\sigma_{z})}{4} dt\bigg).
\end{alignat}
Using \eqref{update1}, we can find the updated state and the final density matrix of the system:
\begin{alignat}{1}
\rho(t+dt)&=|\psi(t+dt)\rangle\langle\psi(t+dt)|\nonumber\\
&=\frac{{\mathcal M}_r\rho (t){\mathcal M}_r^\dagger}{Tr[{\mathcal M}_r \rho (t){\mathcal M}_r^\dagger ]}\nonumber\\
&=\frac{\rho(t)+\iota\Omega_s[\rho(t),\sigma_x]dt-\frac{\iota\sqrt{\alpha}r}{2\tau}[{\mathcal I}-\sigma_z,\rho(t)]dt-\frac{\alpha}{4}\{\rho(t),{\mathcal I}-\sigma_z\}dt}{1-\frac{r^2}{2\tau}dt-\frac{\alpha}{2}dt(1-z)}.
\end{alignat}
Thus, we arrive at the stochastic master equation:
\begin{alignat}{1}
\frac{d\rho}{dt}&=\iota\Omega_s[\rho(t),\sigma_x]-\frac{\iota\sqrt{r \alpha}}{2\tau}[{\mathcal I}-\sigma_z,\rho(t)]-\frac{\alpha}{4}\{\rho(t),{\mathcal I}-\sigma_z\}+\bigg(\frac{r^2}{2\tau}dt+\frac{\alpha}{2}dt(1-z)\bigg)\rho(t)
\end{alignat}
with $\{ \ldots \}$ denoting the anti-commutator. 
\begin{figure*}[htbp!]
    \centering
    \subfloat[]{\includegraphics[width=0.46\textwidth]{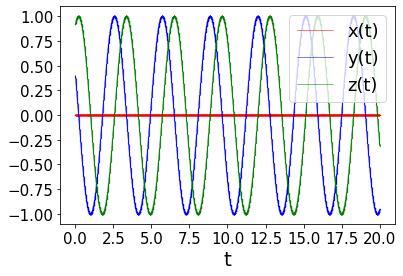}}
    \subfloat[]{\includegraphics[width=0.46\textwidth]{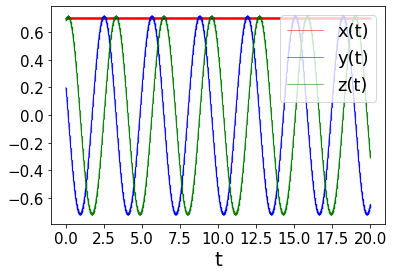}}
   \caption{The evolution of Bloch coordinates without measurement, i.e., $\lambda=0$, for two generic initial conditions, (a) $\{x,y,z\} \equiv \{0,0.4,0.916\}$, $\{p_x,p_y,p_z\}\equiv \{0.5,0.3,0.2\}$, and (b) $\{x,y,z\} \equiv \{0.7,0.2,0.685\}$, $\{p_x,p_y,p_z\}\equiv \{0.2,0.6,0.5\}$. In both the cases, there are periodic Rabi oscillations in the $y$- and $z$-coordinates, whereas the $x$-coordinate remains constant.}
    \label{fig:la=0}
\end{figure*}
Upon comparison of the matrix elements, we arrive at the update equations of Bloch coordinates:
\begin{alignat}{1}
\dot{x}(t)&=-\frac{\alpha x z}{2}+r\sqrt{\frac{\alpha}{\tau}} y, \nonumber\\
\dot{y}(t)&=-\frac{\alpha y z}{2}-r\sqrt{\frac{\alpha}{\tau}} x-2\Omega_s z, \nonumber\\
\dot{z}(t)&=\frac{\alpha(1-z^2)}{2}+2\Omega_s y.
\end{alignat}
The functional $\mathcal{F}$, as described in Sec. \ref{CDJ formalism} is $-\frac{\alpha}{2}(\frac{r^2}{\alpha\tau}+1-z)$. The action ${\mathcal S}$ and stochastic Hamiltonian ${\mathcal H}$ are given by \begin{alignat}{1}
{\mathcal S}& = \int_{0}^{T} dt (-p_x\dot x-p_y\dot y-p_z\dot z-{\mathcal H}),\nonumber\\
{\mathcal H}&=p_x\bigg(-\frac{\alpha x z}{2}+r\sqrt{\frac{\alpha}{\tau}} y\bigg) +p_y\bigg(-\frac{\alpha y z}{2}-r\sqrt{\frac{\alpha}{\tau}} x-2\Omega_s z\bigg)\nonumber\\ &+p_z\bigg(\frac{\alpha}{2}(1-z^2)+2\Omega_s y\bigg)-\frac{\alpha}{2}\bigg(\frac{r^2}{\alpha\tau}+1-z\bigg).
\end{alignat}

\begin{figure*}[htbp!]
    \centering
    \subfloat[]{\includegraphics[width=0.46\textwidth]{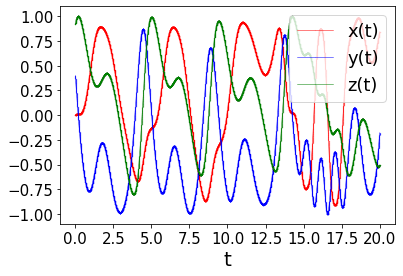}}
    \subfloat[]{\includegraphics[width=0.43\textwidth]{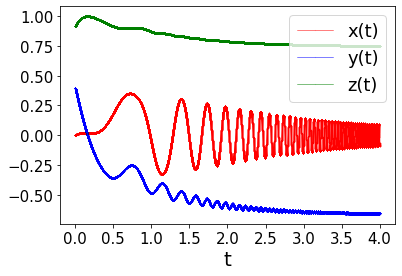}}\\
    \subfloat[]{\includegraphics[width=0.46\textwidth]{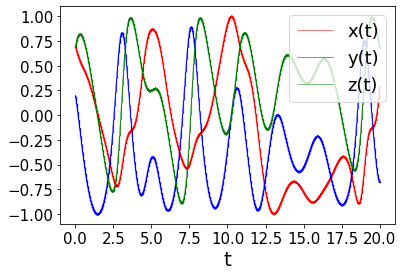}}
    \subfloat[]{\includegraphics[width=0.43\textwidth]{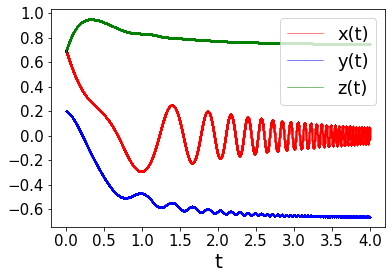}}\\
    \caption{For the same initial conditions as in Fig. \ref{fig:la=0}, (a) and (c) correspond to the case when $\lambda=0.5$, and, (b) and (d) correspond to the case when $\lambda=1.5$. We see that in the Zeno regime ($\lambda=1.5$), the coordinates freeze at the critical point $\{x,y,z\}\equiv\{0,-0.666,0.745\}$, (in spherical coordinates, equivalent to $\theta_1=-\sin^{-1}\left(\frac{1}{1.5}\right)=-0.729$ and $\theta_2=\sin^{-1}\left(\frac{1}{1.5}\right)-\pi=-2.411$ with $\phi_1=\phi_2=\frac{\pi}{2}$, corresponding to the stable and unstable points in Sec. \ref{Critical points}, Fig. \ref{fig:separatrix}).}
    \label{fig:zeno}
\end{figure*}

Extremization of action gives the following coupled differential equations and a constraint on $r$:
\begin{alignat}{1}\label{eq:xyzupdate}
\dot{x}(t)&=-\frac{\alpha x z}{2}+r\sqrt{\frac{\alpha}{\tau}} y,\nonumber\\
\dot{y}(t)&=-\frac{\alpha y z}{2}-r\sqrt{\frac{\alpha}{\tau}} x-2\Omega_s z,\nonumber\\
\dot{z}(t)&=\frac{\alpha}{2}(1-z^2)+2\Omega_s y, \nonumber\\
\dot{p_x}(t)&=\frac{\alpha z p_x}{2}+r\sqrt{\frac{\alpha }{\tau}}p_y,\nonumber\\
\dot{p_y}(t)&=-r\sqrt{\frac{\alpha }{\tau}}p_x+\frac{\alpha z p_y}{2}-2\Omega_s p_z, \nonumber\\
\dot{p_z}(t)&=\frac{\alpha x p_x}{2}+\frac{\alpha y p_y}{2}+2\Omega_s p_y + \alpha z p_z-\frac{\alpha}{2}, \nonumber\\
r&=\sqrt{\alpha\tau}(yp_x-xp_y).
\end{alignat}
In terms of $\theta$ and $\phi$, these critical points correspond to $\theta_1=-\sin^{-1}\left(\frac{1}{1.5}\right)=-0.729$ and $\theta_2=\sin^{-1}\left(\frac{1}{1.5}\right)-\pi=-2.411$ with $\phi_1=\phi_2=\frac{\pi}{2}$. The evolution of density matrix turns out to be stochastic, a consequence of Gaussian-distributed measurement outcomes \cite{Gisint}. The update in Bloch coordinates with time is shown in Fig. \ref{fig:la=0}, \ref{fig:zeno} for different measurement frequencies and different initial conditions. Without measurement, the system exhibits Rabi oscillations (Fig. \ref{fig:la=0}). The measurements influence the dynamics of the Bloch coordinates as shown in Fig. \ref{fig:zeno}. When the measurement frequency is higher than the Rabi frequency, the qubit is frozen in a state (Fig. \ref{fig:zeno} (b), (d)) manifesting the quantum Zeno effect. The phase space point corresponding to this state is the stable point of \eqref{eq:xyzupdate}, indeed equivalent to the stable point $\theta_1$ of Sec. \ref{section 3}.

\section{Concluding Remarks}

In this work, we have employed the action formalism developed by Jordan and coworkers \cite{jordan}  which facilitates a study of evolution of density matrix in time, in terms of evolution of Bloch coordinates and canonically conjugate momenta. The system considered by us has been a subject of close study due to its potential relation with quantum error correction. Unlike in \cite{jordan}, where the measurement readout is performed by a quantum point contact (QPC), we consider another two level system, namely, an ancilla, entangled with our qubit, thus performing a partial measurement rather than a direct one. The dynamics of the system is controlled by the frequency with which repeated measurements are performed leading us to a connection with QZE. For the qubit system considered here, it is shown that the Rabi oscillations are prolonged in time. When the ratio $\lambda > 1$, it has been shown recently that there appear a cascade of stages after the quantum Zeno effect sets in \cite{parveen}. There appear two critical points in $\theta$ - one stable and the other unstable. The system navigates to the stable point in the sense that an ensemble of identically prepared systems evolve towards the stable point. Then, according to the earlier works \cite{parveen}, after some time (duration being statistical), the system would make a transition to the unstable point, from where, it would reach the state $|0\rangle$, owing to inherent quantum fluctuations and tunneling. The mechanism of this rather important dynamics is correctly and completely captured in a description in terms of phase space as we need to consider the evolution of $\theta$ and $p_{\theta}$ to understand the stages. In fact, as shown here, there appear two hyperbolic points in phase space $P_1(\theta_1, p_{\theta_{1}})$ and $P_2(\theta_2, p_{\theta_{2}})$ - the important difference being that $\theta$ ($p_{\theta})$ is a stable (unstable) direction around $P_1$, and, exactly the opposite is the case for $P_2$ \ref{fig:separatrix}. At $P_1$ and $P_2$, the stable and unstable directions are interchanged. Thus, the complete explanation of the stages and the transition is as follows. While the system is attracted towards $P_1$ along $\theta$, due to instability along $p_{\theta}$, it allows the system to reach $P_2$ along $p_{\theta}$ wherefrom the system evolves due to instability along $\theta$ direction. This provides a clear picture of different stages in the QZE as phase space description is complete whereas a description in terms of just the angle or coordinates would be a reduced one. We would also like to point out that the saddle points seen in Fig. \ref{fig:separatrix} are time-reversed partners insofar as $(\theta_1, p_{\theta_1})$ is transformed to the other point via $\pi - \theta_1, -p_{\theta_1}$. So, under time-reversal, the other saddle point plays the role of the state where the system would be attracted to.

Even in the case of diffusive measurement, the system evolves towards the same critical points as in Sec. \ref{section 3}, thus exhibiting quantum Zeno effect. The system demonstrates the lengthening of Rabi period for frequencies of detector lower than the Rabi frequency, and freezing of the coordinates to reach the stable points for frequencies higher than the Rabi frequency, which is reminiscent of the ``critical slowing down", discussed in \cite{parveen}. We would like to conclude by expressing that the usage of the stochastic action principle to analyze the quantum Zeno effect and related qubit dynamics on phase space and Bloch sphere is most insightful. 

\noindent
{\bf Acknowledgement}

We would like to thank Dr. Parveen Kumar for stimulating and instructive discussions.

\end{document}